\documentclass[aps,prl,preprint]{revtex4}
\usepackage{graphicx,booktabs,natbib}
\graphicspath{{./fig/}}
\begin{document}
\title{A radio-frequency sheath model for complex waveforms}
\author{M. M. Turner}
\affiliation{School of Physical Sciences and National Centre for Plasma
Science and Technology, Dublin City University, Dublin 9, Ireland}
\author{P. Chabert}
\affiliation{Laboratoire de Physique des Plasmas, Centre National de
la Recherche Scientifique, Ecole Polytechnique, Universit\'e Pierre
et Marie Curie, Paris XI, 91128 Palaiseau, France}
\begin{abstract}
Plasma sheaths driven by radio-frequency voltages occur frequently, in
contexts ranging from plasma processing applications to magnetically
confined fusion experiments. These sheaths are crucial because they
dominantly affect impedance, power absorption, ion acceleration and
sometimes the stability of the nearby plasma.  An analytical
understanding of sheath behavior is therefore important, both
intrinsically and as an element in more elaborate theoretical
structures.  In practice, these radio-frequency sheaths are commonly excited
by highly anharmonic waveforms, but no analytical model exists for
this general case.  In this letter we present a mathematically simple
sheath model that can be solved for essentially arbitrary excitation
waveforms.  We show that this model is in good agreement with earlier
models for single frequency excitation, and we show by example how to
develop a solution for a complex wave form.  This solution is in good
agreement with simulation data.  This simple and accurate model is
likely to have wide application.
\end{abstract}
\maketitle

In many radio-frequency discharges the sheath is the most important
region, because impedance, power absorption and ion acceleration are
dominated by sheath processes
\cite{lieberman_principles_2005,chabert_physics_2011}.  There are
other contexts where radio-frequency sheath physics is a concern, for
example when understanding the physics of heating in certain fusion
plasmas
\cite{dippolito_radio-frequency-sheath-driven_1993,colas_understanding_2007,myra_resonance_2008}.  Consequently, models of the sheath are important, either in
themselves or as elements in more complex situations, over a broad
area of plasma physics.  The problem has been considered
on a number of occasions, {\em e.g.} \cite{schneider_zum_1954,butler_plasma_1963,godyak_soviet_1986}.  Analytical
models are particularly useful for developing physical insight and
expressing the relationships between parameters in a clear way, but such
models have proved elusive.
Lieberman\cite{lieberman_analytical_1988,lieberman_principles_2005}
supplied an analytical model for a radio-frequency sheath driven by a
single frequency, but in practice much more complex waveforms
frequently occur
\cite{wang_control_2000,heil_possibility_2008,johnson_nanocrystalline_2010,johnson_microcrystalline_2012,lafleur_enhanced_2012}.
There has been limited success in generalizing the Lieberman model to
cover these cases, because of mathematical complexities
\cite{robiche_analytical_2003,franklin_dual_2003,boyle_modelling_2004}.
So there is essentially no sheath model available to describe many
modern experiments.  In this paper we present a new analytical sheath
model, based on a simpler mathematical framework than that of
Lieberman\cite{lieberman_analytical_1988,lieberman_principles_2005}.
For the single frequency case, this model yields scaling laws that are
identical in form to those of
Lieberman\cite{lieberman_analytical_1988,lieberman_principles_2005},
differing only by numerical coefficients close to one.  However, the
new model may be straightforwardly solved for almost arbitrary current
waveforms, and may be used to derive scaling laws for such
cases.

\begin{figure}
\includegraphics[width=0.8\columnwidth]{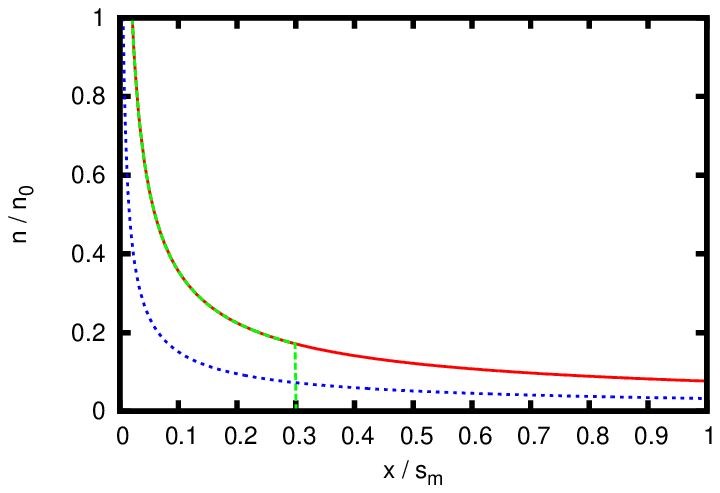}
\includegraphics[width=0.8\columnwidth]{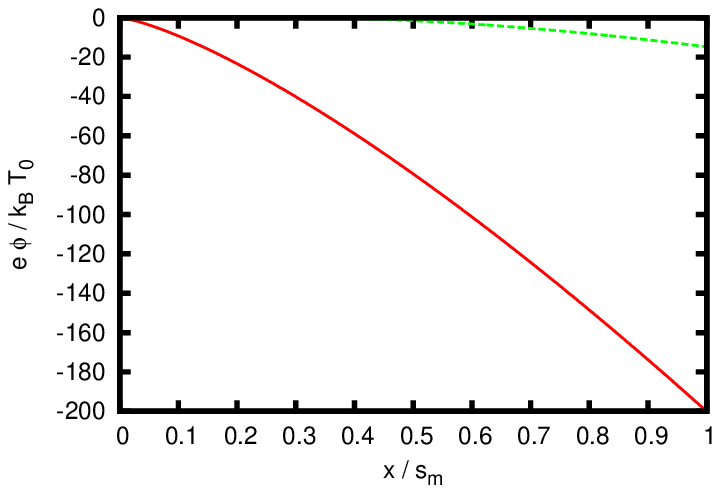}
\caption{Upper panel: Charged particle densities, showing the ion
density, $n_i$ (solid line) the time averaged electron density,
$\bar{n}_e$ (dotted line), and the electron density, $n_e$ at the
instant when $s/s_m=0.3$ (dashed line).  Lower panel: Electrostatic
potential, showing the time averaged potential, $\bar{\phi}$ (solid
line) and the instantaneous potential when $s/s_m=0.3$ (dashed
line).\label{fig_n_phi}}
\end{figure}

Fig.~\ref{fig_n_phi} is a schematic representation of the charged particle
densities and fields that occur in a radio-frequency sheath.
For a sheath in this regime, the ion motion is
determined by the time averaged field, while the electrons respond to
the instantaneous field.  A model describing such a sheath therefore
has a time-averaged part and a time-dependent part, which must be
consistent.  We therefore insist on the same maximum sheath width,
$s_m$, in both cases.  The principal parameter determining $s_m$ in
the time-averaged sense is the time-averaged sheath voltage, $\bar{V}$,
while in the time-dependent model, $s_m$ is a function of $V_0$, the
maximum sheath voltage.  The simplest way to satisfy the constraint
is to choose
\begin{equation}
\frac{\bar{V}}{V_0} = \frac{\bar{\rho}}{\rho_0} \equiv \xi,\label{eq_def_xi}
\end{equation}
where $\bar{\rho}$ is the time-averaged charge density, $\rho_0$ is
the charge density when the sheath voltage is $V_0$, and $\xi$ emerges
as a key parameter of our model.  We note that $\rho_0 \neq
\bar{\rho}$ because of the time dependence of the electron density in
the sheath region.  Since the ion density $n_i$ is time-independent,
these relations imply $\bar{n}_e = (1-\xi) n_i$, which is our central
approximation.  We now consider specifically a sheath adjacent to a
plasma of density $n_0$ and electron temperature $T_0$.  Ions flow
into the sheath at $x=0$ and are absorbed at an electrode at $x=s_m$.
The constant ion current density is $J_i = e n_0 u_B$, where
$u_B=\sqrt{k_B T_0/M}$ and $M$ is the ion mass. Now the governing
equations for the time-averaged ion motion are
\begin{eqnarray}
n_i u_i &=& n_0 u_B \\
e\bar{\phi} + \frac{1}{2} M u_i^2 &=&
    \frac{1}{2}M u_B^2 \approx 0 \\
\frac{d^2\bar{\phi}}{dx^2} &=& \frac{e\left(\bar{n}_e - n_i\right)}{\epsilon_0} = -\frac{e \xi n_i}{\epsilon_0},
\end{eqnarray}
where $u_i$ is the (time independent) ion drift velocity and
$\bar{\phi}$ is the time-averaged potential, and we assume
that $e\bar{V} \gg k_B T_0$.  These equations are
those of the Child-Langmuir sheath model
\cite{child_discharge_1911,langmuir_vapor_1913,lieberman_principles_2005,chabert_physics_2011},
with the addition of the parameter $\xi$, and we can at once write
down the solutions
\begin{eqnarray}
J_i &=& K_i \frac{\epsilon_0}{s_m^2}
     \left(\frac{2e}{M}\right)^\frac{1}{2}
      \left(-\bar{V}\right)^\frac{3}{2}
        \label{eq_child}\\
n_i(x) &=& -\frac{4}{9}\frac{\epsilon_0\bar{V}}{\xi e s_m^2}
        \left(\frac{s_m}{x}\right)^\frac{2}{3}\\
\bar{\phi}(x) &=& \bar{V}\left(\frac{x}{s_m}\right)^\frac{4}{3}\\
\bar{E}(x) &=& -\frac{4}{3}\frac{\bar{V}}{s_m}\left(\frac{x}{s_m}\right)^\frac{1}{3}
\end{eqnarray}
where the boundary condition $\bar{E}(x=0)=0$ has been used, and where $K_i=4/(9\xi)$. The time-dependent
field and potential can now be determined by integrating Poisson's equation
again with the assumption that
\begin{equation}
n_e = \cases{ 0 & if $s < x \le s_m$; \cr n_i & otherwise,\cr}
    \label{eq_step}
\end{equation}
where $s(t)$ is the position of the sheath edge, at which
point $E=0$ and $\phi=0$.  We obtain for $s < x \le s_m$:
\begin{eqnarray}
\phi(x,t) &=& \frac{\bar{V}}{\xi}\left[\left(\frac{x}{s_m}\right)^\frac{4}{3}
              - \frac{4}{3}\left(\frac{s}{s_m}\right)^\frac{1}{3}
                \left(\frac{x}{s_m}\right)
              + \frac{1}{3}\left(\frac{s}{s_m}\right)^\frac{4}{3}\right]\\
E(x,t)    &=& -\frac{4}{3}\frac{\bar{V}}{\xi s_m}\left[
               \left(\frac{x}{s_m}\right)^\frac{1}{3}
              - \left(\frac{s}{s_m}\right)^\frac{1}{3}\right],
\end{eqnarray}
so that the time dependent sheath voltage is
\begin{equation}
V(t) = V_0\left[1
              - \frac{4}{3}\left(\frac{s}{s_m}\right)^\frac{1}{3}
              + \frac{1}{3}\left(\frac{s}{s_m}\right)^\frac{4}{3}\right].
    \label{eq_v}
\end{equation}
Now
\begin{equation}
J = \epsilon_0\left.\frac{\partial E}{\partial t}\right|_{x=s_m}
  = \frac{4}{3}\frac{\epsilon_0 V_0}{s_m}\frac{d}{dt}
    \left(\frac{s}{s_m}\right)^\frac{1}{3},
\end{equation}
so that
\begin{equation}
\frac{s}{s_m} = \left[\frac{3}{4}\frac{s_m}{\epsilon_0 V_0}
               \int_0^t J dt\right]^3.\label{eq_current}
\end{equation}
Since $0 \le s/s_m \le 1$, once $J(t)$ is chosen, $s(t)/s_m$ is fully defined
and we can express
\begin{equation}
\xi = \frac{\left\langle V(t)\right\rangle}{V_0} = \left< 1
              - \frac{4}{3}\left(\frac{s}{s_m}\right)^\frac{1}{3}
              + \frac{1}{3}\left(\frac{s}{s_m}\right)^\frac{4}{3} \right>
    \label{eq_xi_general}.
\end{equation}
Hence, once $J(t)$ has been given, all the remaining quantities can be
calculated without further assumption or approximation, as the
following example shows.

For the single frequency case treated by Lieberman
\cite{lieberman_analytical_1988} we choose $J(t)=-J_0\sin\omega
t$. From Eqs. (\ref{eq_xi_general}) and (\ref{eq_current}) we find
\begin{eqnarray}
s(t) &=& \frac{s_m}{8}\left(1 - \cos\omega t\right)^3\label{eq_s_single}\\
J_0&=&-\frac{K_{\rm cap}}{2}\frac{\omega\epsilon_0}{s_m}V_0\label{eq_j_0_single}\\
\xi&=&\frac{163}{384}
\end{eqnarray}
where $K_{\rm cap}=4/3$. The voltage waveform is obtained by inserting
Eq. (\ref{eq_s_single}) into Eq. (\ref{eq_v}). Combining
Eqs. (\ref{eq_child}) and (\ref{eq_j_0_single}) gives the sheath
maximum expansion as a function of $J_0$,
\begin{equation}
s_m = \frac{K_sJ_0^3}{e\epsilon_0k_BT_0\omega^3n_0^2},\label{eq_s_m_single}
\end{equation}
where $K_s=4\xi/3$. These expressions are identical with those of Lieberman
\cite{lieberman_analytical_1988,lieberman_principles_2005}, apart from
numerical coefficients close to unity.  Table~\ref{tab_coefficients}
compares $\xi$, $K_i$, $K_{\rm cap}$ and $K_s$ for the two models and
shows that they are not significantly different.
Similarly, the
voltage waveforms are almost identical in the two models. 
 This may seem surprising, in view of the apparently bold approximation
of Eq.~(\ref{eq_def_xi}).
However, \citet{brinkmann_electron_2009} has shown that the
approximation of Eq.~(\ref{eq_step}) is important, and because this
approximation is integral to the Lieberman model, the accuracy of that
model is not significantly better than the present one.  However, the
present model can be solved for a far greater range of waveforms.  In
particular, the current density can be expressed as an arbitrary
Fourier series, leading, for example, to models for multiple-frequency
excitation that are free of inconvenient restrictions on the component
amplitudes (such as occur in dual-frequency generalizations of
Lieberman's model
\cite{robiche_analytical_2003,franklin_dual_2003,boyle_modelling_2004}).

For example, we can choose
\begin{equation}
J(t) = -J_0\sin\omega_0 t - J_1\sin\omega_1t
\end{equation}
and find at once
\begin{eqnarray}
s(t) &=&
    \frac{s_m}{8}\left[\frac{(J_0/\omega_0)(1-\cos\omega_0t)
        + (J_1/\omega_1)(1-\cos\omega_1t)}{J_0/\omega_0 + J_1/\omega_1}\right]^3
    \\
J_0&=&-\frac{2}{3}\frac{\omega_0\epsilon_0V_0}{s_m}\left(1+\frac{J_1\omega_0}{J_0\omega_1}\right)\\
s_m &=& \frac{4\xi}{3}\frac{\left(J_0/\omega_0 + J_1/\omega_1\right)^3}
                   {\epsilon_0 e n_0^2 k_B T_0}\\
\xi &=& \frac{1}{3} + \frac{\frac{35}{128}
                      \left[\left(\frac{J_0}{\omega_0}\right)^4
                           + \left(\frac{J_1}{\omega_1}\right)^4\right]
                    + \frac{5}{8}\left[\left(\frac{J_0}{\omega_0}\right)^3
                      \frac{J_1}{\omega_1}
                      + \frac{J_0}{\omega_0}
                      \left(\frac{J_1}{\omega_1}\right)^3\right]
                    + \frac{27}{32}\left(\frac{J_0}{\omega_0}\right)^2
                      \left(\frac{J_1}{\omega_1}\right)^2}
                    {\left(J_0/\omega_0 + J_1/\omega_1\right)^4}
                    \label{eq_xi_dual}\\
    &\approx& \frac{163}{384},
\end{eqnarray}
where we note that eq.~\ref{eq_xi_dual} yields a result never different
by more than 10~\% from the single frequency result, which value can therefore
be used for all practical purposes.
These formulae are therefore more generally applicable, less cumbersome
and obtained with less mathematical exertion than those previously
given \cite{robiche_analytical_2003,franklin_dual_2003,boyle_modelling_2004}.

\begin{figure}
\includegraphics[width=0.8\columnwidth]{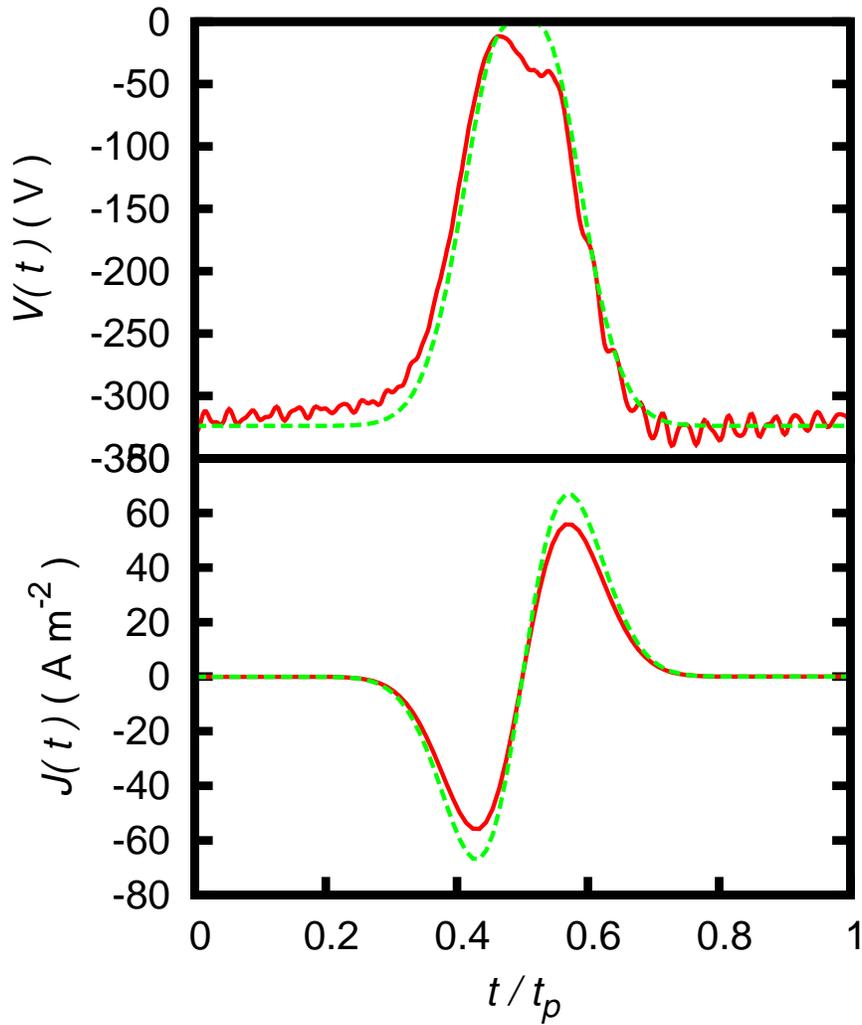}
\caption{Comparison of simulation results (solid lines) with
the analytical theory of the text (dashed lines) for sheath voltage
(upper panel) and sheath current density (lower panel).  For this
case $t_w=5.2$~ns.  The electrical control parameter for the
sheath model is $\bar{V}$, the time averaged sheath voltage, which
is here chosen to be the same as in the simulation.\label{fig_j_v}}
\end{figure}

\begin{figure}
\includegraphics[width=\columnwidth]{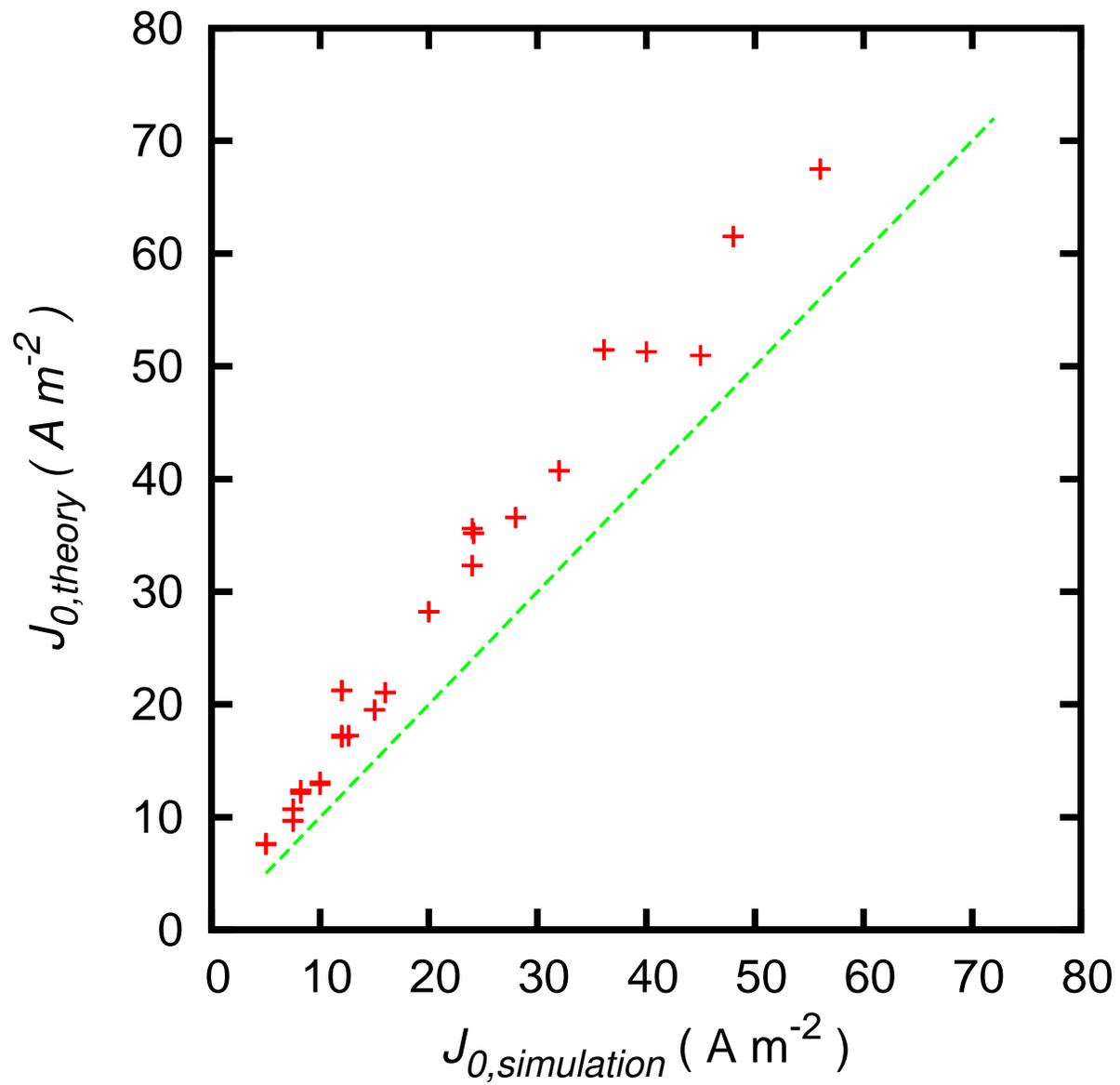}
\caption{Comparison of the maximum radio-frequency current density
found in simulation (horizontal axis) with the result computed from
eq.~\ref{eq_j0_pulse} (vertical axis).  The solid line denotes ideal
agreement between theory and simulation.
\label{fig_j0}}
\end{figure}

\begin{figure}
\includegraphics[width=\columnwidth]{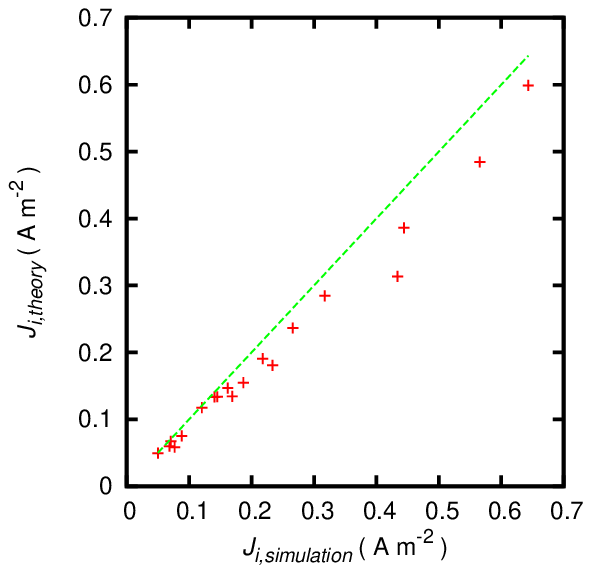}
\caption{Comparison of the ion current density found in simulation
(horizontal axis) with the result computed from eq.~\ref{eq_child}
and eq.~\ref{eq_xi_pulse}
(vertical axis).  The solid line denotes ideal agreement between theory
and simulation.
\label{fig_child}}
\end{figure}

As a third example
we consider a sheath excited by the pulsed waveform
\begin{equation}
J(t) = J_0\left(\frac{t}{t_w}\right)
       \exp\left(\frac{1}{2}-\frac{1}{2}\frac{t^2}{t_w^2}\right),
       \label{eq_pulse}
\end{equation}
which is representative of several topical experiments
\cite{wang_control_2000,heil_possibility_2008,johnson_nanocrystalline_2010}.
We assume that this pulse is repeated at intervals $t_p\ll t_w$, such
that successive pulses do not appreciably overlap.  In this case we
find
\begin{eqnarray}
s(t) &=& s_m\exp\left(-\frac{3}{2}\frac{t^2}{t_w^2}\right)\\
\xi &=& 1 - \frac{7}{3}\sqrt{\frac{\pi}{2}}\frac{t_w}{t_p}\label{eq_xi_pulse}\\
J_0 &=& -\frac{4}{3} \frac{\epsilon_0 V_0}{s_m t_w \exp\left(\frac{1}{2}\right)}
     \label{eq_j0_pulse}\\
s_m &=& \frac{\xi}{6} \exp\left(\frac{3}{2}\right)
        \frac{\left(J_0 t_w\right)^3}{\epsilon_0 e n_0^2 k_B T_0}
\end{eqnarray}
The Child law, and therefore the sheath width, are found by inserting
Eq.~(\ref{eq_xi_pulse}) into Eq.~(\ref{eq_child}).  We have investigated the utility of this model
by comparison with particle-in-cell simulation data
\cite{birdsall_plasma_1991,birdsall_particle--cell_1991}.  These
simulations treated a plasma formed in a space between two plane
parallel electrodes separated by 6.7~cm, filled with argon gas at a
pressure of 10~mTorr, and excited by a current density with the form
of Eq.~(\ref{eq_pulse}).  The peak current density ranged from 5 to 70
A~m$^{-2}$ and the pulse width from approximately 1 to 10~ns.  These
conditions lead to $T_0\approx 1.5$~eV and $n_0 \approx 3\times
10^{14}$--$3\times 10^{15}$~m$^{-3}$. In figs.~\ref{fig_j_v},
\ref{fig_j0} and \ref{fig_child} we compare these simulation results
with the predictions of the present model, and we find good agreement,
both for time time dependent currents and voltages, and scaling laws.

To summarize, in this paper we have developed a new sheath model that can be
expressed in a small number of straightforward equations.  For the
single-frequency case, the model agrees well with the Lieberman model.
The Lieberman model, in spite of its elegant construction, is
mathematically complex and has proved resistant to generalization. The
present model, however, is readily adaptable to a wide range of
complex excitation waveforms, and leads to results in good agreement
with simulations.  Thus this model provides a valuable tool for designing
and understanding experiments involving non-sinusoidal excitation
of radio-frequency sheaths.

\begin{acknowledgments}
The work of MMT was supported by Science Foundation Ireland under grant numbers
07/IN.1/I907 and 08/SRC/I1411.  The work of PC was supported by the Agence
Nationale de le Recherche (CANASTA Project No.\ ANR-10-HABISOL-002).
\end{acknowledgments}

\begin{table}
\begin{center}
\begin{tabular}{cccc}
\hline
      & Present & Lieberman  \\
\hline
$\xi$ & 0.425    & 0.415 \\
$K_{\rm cap}$ & 1.33    & 1.23 \\
$K_i$ & 1.05    & 0.82 \\
$K_s$ & 0.566 & 0.417 \\
\bottomrule
\end{tabular}
\caption{Comparison of numerical coefficients in eqs.~\ref{eq_s_m_single},
\ref{eq_j_0_single} and \ref{eq_child} determined from three
different models.\label{tab_coefficients}}
\end{center}
\end{table}

\begin{thebibliography}{22}
\expandafter\ifx\csname natexlab\endcsname\relax\def\natexlab#1{#1}\fi
\expandafter\ifx\csname bibnamefont\endcsname\relax
  \def\bibnamefont#1{#1}\fi
\expandafter\ifx\csname bibfnamefont\endcsname\relax
  \def\bibfnamefont#1{#1}\fi
\expandafter\ifx\csname citenamefont\endcsname\relax
  \def\citenamefont#1{#1}\fi
\expandafter\ifx\csname url\endcsname\relax
  \def\url#1{\texttt{#1}}\fi
\expandafter\ifx\csname urlprefix\endcsname\relax\def\urlprefix{URL }\fi
\providecommand{\bibinfo}[2]{#2}
\providecommand{\eprint}[2][]{\url{#2}}

\bibitem[{\citenamefont{Lieberman and
  Lichtenberg}(2005)}]{lieberman_principles_2005}
\bibinfo{author}{\bibfnamefont{M.~A.} \bibnamefont{Lieberman}}
  \bibnamefont{and} \bibinfo{author}{\bibfnamefont{A.~J.}
  \bibnamefont{Lichtenberg}}, \emph{\bibinfo{title}{Principles Of Plasma
  Discharges and Materials Processing}} (\bibinfo{publisher}{John Wiley \&
  Sons}, \bibinfo{year}{2005}).

\bibitem[{\citenamefont{Chabert and Braithwaite}(2011)}]{chabert_physics_2011}
\bibinfo{author}{\bibfnamefont{P.}~\bibnamefont{Chabert}} \bibnamefont{and}
  \bibinfo{author}{\bibfnamefont{N.~S.~J.} \bibnamefont{Braithwaite}},
  \emph{\bibinfo{title}{Physics of Radio-Frequency Plasmas}}
  (\bibinfo{publisher}{Cambridge University Press}, \bibinfo{year}{2011}).

\bibitem[{\citenamefont{{D{\textquoteright}Ippolito}
  et~al.}(1993)\citenamefont{{D{\textquoteright}Ippolito}, Myra, Jacquinot, and
  Bures}}]{dippolito_radio-frequency-sheath-driven_1993}
\bibinfo{author}{\bibfnamefont{D.~A.}
  \bibnamefont{{D{\textquoteright}Ippolito}}},
  \bibinfo{author}{\bibfnamefont{J.~R.} \bibnamefont{Myra}},
  \bibinfo{author}{\bibfnamefont{J.}~\bibnamefont{Jacquinot}},
  \bibnamefont{and} \bibinfo{author}{\bibfnamefont{M.}~\bibnamefont{Bures}},
  \bibinfo{journal}{Physics of Fluids B: Plasma Physics}
  \textbf{\bibinfo{volume}{5}}, \bibinfo{pages}{3603} (\bibinfo{year}{1993}).

\bibitem[{\citenamefont{Colas et~al.}(2007)\citenamefont{Colas, Ekedahl,
  Goniche, Gunn, Nold, Corre, Bobkov, Dux, Braun, Noterdaeme
  et~al.}}]{colas_understanding_2007}
\bibinfo{author}{\bibfnamefont{L.}~\bibnamefont{Colas}},
  \bibinfo{author}{\bibfnamefont{A.}~\bibnamefont{Ekedahl}},
  \bibinfo{author}{\bibfnamefont{M.}~\bibnamefont{Goniche}},
  \bibinfo{author}{\bibfnamefont{J.~P.} \bibnamefont{Gunn}},
  \bibinfo{author}{\bibfnamefont{B.}~\bibnamefont{Nold}},
  \bibinfo{author}{\bibfnamefont{Y.}~\bibnamefont{Corre}},
  \bibinfo{author}{\bibfnamefont{V.}~\bibnamefont{Bobkov}},
  \bibinfo{author}{\bibfnamefont{R.}~\bibnamefont{Dux}},
  \bibinfo{author}{\bibfnamefont{F.}~\bibnamefont{Braun}},
  \bibinfo{author}{\bibfnamefont{J.-M.} \bibnamefont{Noterdaeme}},
  \bibnamefont{et~al.}, \bibinfo{journal}{Plasma Physics and Controlled Fusion}
  \textbf{\bibinfo{volume}{49}}, \bibinfo{pages}{B35} (\bibinfo{year}{2007}).

\bibitem[{\citenamefont{Myra and
  {D{\textquoteright}Ippolito}}(2008)}]{myra_resonance_2008}
\bibinfo{author}{\bibfnamefont{J. R.}~\bibnamefont{Myra}} \bibnamefont{and}
  \bibinfo{author}{\bibfnamefont{D. A.}~\bibnamefont{{D{\textquoteright}Ippolito}%
}}, \bibinfo{journal}{Physical Review Letters} \textbf{\bibinfo{volume}{101}},
  \bibinfo{pages}{195004}
  (\bibinfo{year}{2008}).

\bibitem[{\citenamefont{Schneider}(1954)}]{schneider_zum_1954}
\bibinfo{author}{\bibfnamefont{F.}~\bibnamefont{Schneider}},
  \bibinfo{journal}{Zeitschrift fur angewandte Physik}
  \textbf{\bibinfo{volume}{6}}, \bibinfo{pages}{456} (\bibinfo{year}{1954}).

\bibitem[{\citenamefont{Butler and Kino}(1963)}]{butler_plasma_1963}
\bibinfo{author}{\bibfnamefont{H.~S.} \bibnamefont{Butler}} \bibnamefont{and}
  \bibinfo{author}{\bibfnamefont{G.~S.} \bibnamefont{Kino}},
  \bibinfo{journal}{Physics of Fluids} \textbf{\bibinfo{volume}{6}},
  \bibinfo{pages}{1346} (\bibinfo{year}{1963}).

\bibitem[{\citenamefont{Godyak}(1986)}]{godyak_soviet_1986}
\bibinfo{author}{\bibfnamefont{V.~A.} \bibnamefont{Godyak}},
  \emph{\bibinfo{title}{Soviet Radio Frequency Discharge Research}}
  (\bibinfo{publisher}{Delphic Associates}, \bibinfo{year}{1986}).

\bibitem[{\citenamefont{Lieberman}(1988)}]{lieberman_analytical_1988}
\bibinfo{author}{\bibfnamefont{M.}~\bibnamefont{Lieberman}},
  \bibinfo{journal}{Plasma Science, {IEEE} Transactions on}
  \textbf{\bibinfo{volume}{16}}, \bibinfo{pages}{638 } (\bibinfo{year}{1988}).

\bibitem[{\citenamefont{Wang and Wendt}(2000)}]{wang_control_2000}
\bibinfo{author}{\bibfnamefont{S.-B.} \bibnamefont{Wang}} \bibnamefont{and}
  \bibinfo{author}{\bibfnamefont{A.~E.} \bibnamefont{Wendt}},
  \bibinfo{journal}{Journal of Applied Physics} \textbf{\bibinfo{volume}{88}},
  \bibinfo{pages}{643} (\bibinfo{year}{2000}).

\bibitem[{\citenamefont{Heil et~al.}(2008)\citenamefont{Heil, Czarnetzki,
  Brinkmann, and Mussenbrock}}]{heil_possibility_2008}
\bibinfo{author}{\bibfnamefont{B.~G.} \bibnamefont{Heil}},
  \bibinfo{author}{\bibfnamefont{U.}~\bibnamefont{Czarnetzki}},
  \bibinfo{author}{\bibfnamefont{R.~P.} \bibnamefont{Brinkmann}},
  \bibnamefont{and}
  \bibinfo{author}{\bibfnamefont{T.}~\bibnamefont{Mussenbrock}},
  \bibinfo{journal}{Journal of Physics D: Applied Physics}
  \textbf{\bibinfo{volume}{41}}, \bibinfo{pages}{165202}
  (\bibinfo{year}{2008}).

\bibitem[{\citenamefont{Johnson et~al.}(2010)\citenamefont{Johnson, Verbeke,
  Vanel, and Booth}}]{johnson_nanocrystalline_2010}
\bibinfo{author}{\bibfnamefont{E.~V.} \bibnamefont{Johnson}},
  \bibinfo{author}{\bibfnamefont{T.}~\bibnamefont{Verbeke}},
  \bibinfo{author}{\bibfnamefont{J.-C.} \bibnamefont{Vanel}}, \bibnamefont{and}
  \bibinfo{author}{\bibfnamefont{J.-P.} \bibnamefont{Booth}},
  \bibinfo{journal}{Journal of Physics D: Applied Physics}
  \textbf{\bibinfo{volume}{43}}, \bibinfo{pages}{412001}
  (\bibinfo{year}{2010}).

\bibitem[{\citenamefont{Johnson et~al.}(2012)\citenamefont{Johnson, Delattre,
  and Booth}}]{johnson_microcrystalline_2012}
\bibinfo{author}{\bibfnamefont{E.~V.} \bibnamefont{Johnson}},
  \bibinfo{author}{\bibfnamefont{P.~A.} \bibnamefont{Delattre}},
  \bibnamefont{and} \bibinfo{author}{\bibfnamefont{J.~P.} \bibnamefont{Booth}},
  \bibinfo{journal}{Applied Physics Letters} \textbf{\bibinfo{volume}{100}},
  \bibinfo{pages}{133504} (\bibinfo{year}{2012}).

\bibitem[{\citenamefont{Lafleur et~al.}(2012)\citenamefont{Lafleur, Boswell,
  and Booth}}]{lafleur_enhanced_2012}
\bibinfo{author}{\bibfnamefont{T.}~\bibnamefont{Lafleur}},
  \bibinfo{author}{\bibfnamefont{R.~W.} \bibnamefont{Boswell}},
  \bibnamefont{and} \bibinfo{author}{\bibfnamefont{J.~P.} \bibnamefont{Booth}},
  \bibinfo{journal}{Applied Physics Letters} \textbf{\bibinfo{volume}{100}},
  \bibinfo{pages}{194101} (\bibinfo{year}{2012}).

\bibitem[{\citenamefont{Robiche et~al.}(2003)\citenamefont{Robiche, Boyle,
  Turner, and Ellingboe}}]{robiche_analytical_2003}
\bibinfo{author}{\bibfnamefont{J.}~\bibnamefont{Robiche}},
  \bibinfo{author}{\bibfnamefont{P.~C.} \bibnamefont{Boyle}},
  \bibinfo{author}{\bibfnamefont{M.~M.} \bibnamefont{Turner}},
  \bibnamefont{and} \bibinfo{author}{\bibfnamefont{A.~R.}
  \bibnamefont{Ellingboe}}, \bibinfo{journal}{Journal of Physics D: Applied
  Physics} \textbf{\bibinfo{volume}{36}}, \bibinfo{pages}{1810}
  (\bibinfo{year}{2003}).

\bibitem[{\citenamefont{Franklin}(2003)}]{franklin_dual_2003}
\bibinfo{author}{\bibfnamefont{R.~N.} \bibnamefont{Franklin}},
  \bibinfo{journal}{Journal of Physics D: Applied Physics}
  \textbf{\bibinfo{volume}{36}}, \bibinfo{pages}{2660} (\bibinfo{year}{2003}).

\bibitem[{\citenamefont{Boyle et~al.}(2004)\citenamefont{Boyle, Robiche, and
  Turner}}]{boyle_modelling_2004}
\bibinfo{author}{\bibfnamefont{P.~C.} \bibnamefont{Boyle}},
  \bibinfo{author}{\bibfnamefont{J.}~\bibnamefont{Robiche}}, \bibnamefont{and}
  \bibinfo{author}{\bibfnamefont{M.~M.} \bibnamefont{Turner}},
  \bibinfo{journal}{Journal of Physics D: Applied Physics}
  \textbf{\bibinfo{volume}{37}}, \bibinfo{pages}{1451} (\bibinfo{year}{2004}).

\bibitem[{\citenamefont{Child}(1911)}]{child_discharge_1911}
\bibinfo{author}{\bibfnamefont{C.~D.} \bibnamefont{Child}},
  \bibinfo{journal}{Physical Review {(Series} I)}
  \textbf{\bibinfo{volume}{32}}, \bibinfo{pages}{492} (\bibinfo{year}{1911}).

\bibitem[{\citenamefont{Langmuir}(1913)}]{langmuir_vapor_1913}
\bibinfo{author}{\bibfnamefont{I.}~\bibnamefont{Langmuir}},
  \bibinfo{journal}{Physical Review} \textbf{\bibinfo{volume}{2}},
  \bibinfo{pages}{329} (\bibinfo{year}{1913}).

\bibitem[{\citenamefont{Brinkmann}(2009)}]{brinkmann_electron_2009}
\bibinfo{author}{\bibfnamefont{R.~P.} \bibnamefont{Brinkmann}},
  \bibinfo{journal}{Journal of Physics D: Applied Physics}
  \textbf{\bibinfo{volume}{42}}, \bibinfo{pages}{194009}
  (\bibinfo{year}{2009}).

\bibitem[{\citenamefont{Birdsall and Langdon}(1991)}]{birdsall_plasma_1991}
\bibinfo{author}{\bibfnamefont{C.~K.} \bibnamefont{Birdsall}} \bibnamefont{and}
  \bibinfo{author}{\bibfnamefont{A.~B.} \bibnamefont{Langdon}},
  \emph{\bibinfo{title}{Plasma physics via computer simulation}}
  (\bibinfo{publisher}{Adam Hilger}, \bibinfo{year}{1991}).

\bibitem[{\citenamefont{Birdsall}(1991)}]{birdsall_particle--cell_1991}
\bibinfo{author}{\bibfnamefont{C.~K.} \bibnamefont{Birdsall}},
  \bibinfo{journal}{Plasma Science, {IEEE} Transactions on}
  \textbf{\bibinfo{volume}{19}}, \bibinfo{pages}{65 } (\bibinfo{year}{1991}).

\end{thebibliography}
\end{document}